\documentclass[twocolumn,showpacs,superscriptaddress,preprintnumbers,amsmath,amssymb,footinbib]{revtex4}

\usepackage{graphicx}
\usepackage{dcolumn}
\usepackage{bm}

\newcommand{\bq}{\begin{equation}}
\newcommand{\eq}{\end{equation}}
\newcommand{\bqa}{\begin{eqnarray}}
\newcommand{\eqa}{\end{eqnarray}}

\newcommand{\del}{\partial}

\newcommand{\dd}{{\rm d}}

\def\be     {\begin{equation}}
\def\ee     {\end{equation}}
\def\bea        {\begin{eqnarray}}
\def\eea        {\end{eqnarray}}
\def\bnn    {\begin{eqnarray*}}
\def\enn    {\end{eqnarray*}}

\makeatletter
\def\eqnarray{ \stepcounter{equation} \let\@currentlabel=\theequation
\global\@eqnswtrue
\global\@eqcnt\z@
\tabskip\@centering
\let\\=\@eqncr
$$\halign to \displaywidth\bgroup\@eqnsel\hskip\@centering
$\displaystyle\tabskip\z@{##}$&\global\@eqcnt\@ne
\hfil$\displaystyle{{}##{}}$\hfil
&\global\@eqcnt\tw@$\displaystyle\tabskip\z@{##}$\hfil
\tabskip\@centering&\llap{##}\tabskip\z@\cr}
\makeatother

\makeatletter
\def\@arrayacol{\edef\@preamble{\@preamble \hskip .5\arraycolsep}}
\def\array{\let\@acol\@arrayacol \let\@classz\@arrayclassz
\let\@classiv\@arrayclassiv \let\\\@arraycr\def\@halignto{}\@tabarray}
\makeatother
\renewcommand{\arraystretch}{1.6}
\makeatletter
\newcounter{subeqncnt}
\def\thesubeqncnt{\alph{subeqncnt}}
\def\subequations{\begingroup%
\stepcounter{equation}\edef\@tempa{\theequation}%
\let\c@equation\c@subeqncnt\c@subeqncnt\z@
\edef\theequation{\@tempa\noexpand\thesubeqncnt}}

\makeatother
\begin{document}

\title{Interplay between interaction and chiral anomaly 
in the holographic approach}

\author{Ki-Seok Kim$^{1,2}$ and Takuya Tsukioka }

\affiliation{
Asia Pacific Center for Theoretical Physics, POSTECH,
Pohang, Gyeongbuk 790-784, Korea \\ 
$^2$ Department of Physics,
POSTECH, Pohang, Gyeongbuk 790-784, Korea, \\
{\sf kimks, tsukioka@apctp.org}
}

\preprint{\tt APCTP-Pre2011-013}
\preprint{\tt arXiv:1106.0398[cond-mat.str-el]}

\begin{abstract}
Strongly coupled conformal field theory appears to describe
universal scaling around quantum criticality, where critical
exponents reflect the nature of emergent excitations. In
particular, novel symmetries can emerge from strong interactions,
expected to be responsible for quantum number fractionalization.
The underlying mechanism has been proposed that an emergent
enhanced symmetry allows a topological term associated with
anomaly, which assigns a fermion's quantum number to a topological
excitation, referred as the Goldstone-Wilczek current. Although
this mechanism has been verified in one dimensional interacting
electrons, where either spinons or holons are identified with
topological solitons, its generalization to higher dimensions is
beyond the field theoretic framework in the respect that interplay
between interaction and anomaly cannot be taken into account
sincerely within the field theory technique above one dimension.
In this study we examine the interplay between correlations and
topological terms based on the holographic approach, allowing us
to incorporate such nonperturbative quantum effects via solving
classical equations of motion but on a curved space. We solve the
Einstein-Maxwell-Chern-Simons theory on the
Reissner-Nordstr\"om-AdS$_5$ in the extremal limit, and uncover
novel critical exponents to appear in the current-current
correlation functions, where both the emergent locality and the
Chern-Simons term play an important role in such critical
exponents. We speculate that the corresponding conformal field
theory may result from interacting $U(1)$ currents with chiral
anomaly at finite density, expected to be applicable to
topological insulators with strong interactions, where dyon-type
excitations appear to carry nontrivial fermion's quantum numbers.
\end{abstract}

\pacs{11.25.Tq, 71.10.Hf} 
%11.25.Tq 	Gauge/string duality 
%71.10.Hf 	Non-Fermi-liquid ground states, electron phase diagrams and phase transitions in model systems 

\maketitle

\section{introduction}

Universal scaling is an essential aspect in critical phenomena,
where critical exponents imply the information of emergent
excitations~\cite{Scaling_ReviewI,Scaling_ReviewII}. 
Elementary excitations become strongly coupled in the vicinity of quantum
criticality, described by conformal field theories (CFTs), which
do not allow well defined excitations~\cite{QPT_ReviewI,QPT_ReviewII}. 
One can consider two possible scenarios for disappearance of
quasiparticles. 
One immediately suggests that elementary excitations decay into 
bunch of soft modes, given by composite particles in terms of original
excitations. 
This corresponds to usual situations in critical phenomena~\cite{Ma_Book}. 
On the other hand, elementary excitations may become fractionalized 
into more ``basic'' ingredients, which also leads the quasiparticle weight to vanish.
When quantum number fractionalization occurs, critical exponents
are enhanced, originating from the fact that elementary
excitations are given by composite particles in terms of such
fractionalized excitations~\cite{Senthil_DQCP}.

The second scenario is actually well known in one dimensional
correlated systems, where such fractionalized excitations are
referred as spinons and holons~\cite{Tsvelik_Book}. The underlying
mechanism has been proposed that novel symmetries can emerge from
strong interactions, expected to be responsible for novel quantum
numbers. The emergent enhanced symmetry allows a topological term
associated with anomaly, which assigns a fermion's quantum number
to a topological excitation, referred as the Goldstone-Wilczek
current~\cite{GW_Current}. However, its generalization to higher
dimensions~\cite{Senthil_DQCP} is beyond the field theoretic
framework because interplay between interaction and anomaly cannot
be taken into account sincerely within the field theory technique
above one dimension. Only several simulation results have been
reported in extended Heisenberg models~\cite{Sandvik} and some
statistical field theories~\cite{Balents}.

In string theory, it has been clarified that strongly coupled CFTs in
$d$-dimensions can be mapped into classical gravity theories on
Anti-de Sitter space in $(d+1)$-dimensions
(AdS$_{d+1}$)~\cite{AdS_CFT_Conjecture1,AdS_CFT_Conjecture2}. This
duality may be regarded as generalization of the well known
holography of the bulk-edge correspondence in the quantum Hall
effect, where this framework has been developed in the context of
string theory, referred as the AdS/CFT correspondence. See
Ref.~\cite{AdS_CFT_Review} for a review. Immediately, it has been
applied to various problems beyond techniques of field theories:
nonperturbative phenomena in quantum chromodynamics (AdS/QCD or
holographic QCD) \cite{hQCD_review}, non-Fermi liquid transport
near quantum criticality
\cite{AdS_CMP_TR1,AdS_CMP_TR2,AdS_CMP_TR3} and superconductors
\cite{H sconductor gubser,H superconductor
HHH,H_superconductor_Kim} in condensed matter physics (AdS/CMP),
and etc.

The motivation for the usage of the AdS/CFT machinery lies in
solving two difficult problems beyond the field theoretical
framework. Although vector models are expected to be under control
generally, interacting fermions at finite density, named as the
Fermi surface problem, turns out to be out of control, which
displays essentially the same aspect as the matrix model, where
the infinite limit of the flavor degeneracy is impossible to be
performed within the present technology because all planar
diagrams are exactly at the same order and they all should be
summed
\cite{N_infinite_SSL,N_infinite_Sachdev1,
N_infinite_Sachdev2,N_infinite_Senthil, N_infinite_KSKIM}.
In addition to the treatment of strong correlations, we do not
know how to take into account the topological term
nonperturbatively above one dimension, as discussed before. The
AdS/CFT duality tells us that we can incorporate both strong
correlations and topological effects just via solving classical
equations of motion, but in one dimension higher than that of the
field theory and on the curved space.

In this study we examine the interplay between correlations and
topological terms based on the holographic approach. We solve the
Einstein-Maxwell-Chern-Simons (EM-CS) theory on the
Reissner-Nordstr\"om-AdS$_{5}$ (RN-AdS$_5$) 
in the extremal limit, and uncover
novel critical exponents to appear in the current-current
correlation functions, where both the emergent locality and the
Chern-Simons term play an important role in such critical
exponents.
We speculate that the corresponding conformal field theory may
result from strongly interacting $U(1)$ charge currents at finite
density in one time and three space dimensions. In particular,
such a field theory is expected to contain the topological
$\theta$ term, associated with anomalous ``chiral'' currents. We
suggest that strong correlations in the presence of $\theta$ vacua
may result in novel emergent excitations, reflected in their
critical exponents and distinguished from ``boring'' excitations in
the absence of the $\theta$ term. We discuss that such a conformal
field theory may appear at quantum criticality in topological
insulators with fractional magnetoelectric effect
\cite{FTI_EFT_Theta}. However, we cannot interpret the emergence
of locality in the field theoretic point of view, which occurs in
the case when the dynamical critical exponent becomes infinite.
Although it was demonstrated that the dynamical critical exponent
can be changed due to higher-loop quantum corrections
\cite{N_infinite_Sachdev2}, the infinite dynamical exponent does
not seem to be reachable within the present technology except for
the disorder-driven Anderson localization
\cite{Sachdev_CFT}\footnote{
%%% footnote %%%
Although the term ``Anderson localization'' 
itself already contains ``disorder-driven'', 
we use such a repeated expression in order to make not condensed-matter
physics people understandable. 
}. 

\section{Einstein-Maxwell-Chern-Simons} 
 
We start by giving 5D gravity side description 
which may be dual to 4D field theory. 
In addition to the usual Einstein gravity, 
we consider the 5D Maxwell field which could couples to the 
$U(1)$ current on the boundary. 
We can also introduce the CS term in the 5D spacetime. 
The action which we would like to work with is the 
EM-CS in 5D   
with the gravitational constant $G_5$, 
the negative cosmological constant $\Lambda(=-6/l^2)$, the gauge 
coupling $e^2$ and the Chern-Simons coupling $\kappa$: 
\begin{equation}
S=S_{\rm EH}+S_{\rm GH}+S_{\rm Maxwell}+S_{\rm CS},
\label{EMCS}
\end{equation}
where
\begin{subequations}
\begin{eqnarray}
S_{\rm EH}
&=&
\frac{1}{16\pi G_5} \int\!\dd^5x\sqrt{-g}
\big( R -2\Lambda \big), 
\label{eh}
\\
S_{\rm GH}
&=&
\frac{1}{8\pi G_5}
\!\int\!
\dd^4x\sqrt{-g^{(4)}}K,
\label{gh}
\\
S_{\rm Maxwell}
&=& 
- \frac{1}{4e^2}\int\!\dd^5x \sqrt{-g} {\cal F}_{mn} {\cal F}^{mn},
\label{maxwell}
\\
S_{\rm CS}
&=&
\frac{\kappa}{3} \int\!\dd^5x \ \varepsilon^{lmnpq}
{\cal A}_l {\cal F}_{mn}{\cal F}_{pq}.
\label{cs}
\end{eqnarray}
\end{subequations}

\noindent
In order to compensate well-defined variational principle for
the Einstein-Hilbert action $S_{\rm EH}$ with boundary,
we have introduced the Gibbons-Hawking term $S_{\rm GH}$
in which $g^{(4)}_{\mu\nu}$ and $K$ are the induced metric and the
extrinsic curvature on the 4D boundary, respectively.
Throughout of this paper, 
we follow the notation given in the previous work~\cite{mstt}.  

\subsection{RN-AdS$_{\bm5}$ background} 

Equations of motion read
\begin{subequations}
\begin{eqnarray}
R_{mn}-\frac{1}{2}g_{mn}R  -\frac{6}{l^2}g_{mn}
&=&
8\pi G_5 T_{mn}, 
\qquad 
\label{eom_g}
\\
-\frac{1}{e^2} \nabla_n {\cal F}^{mn}+\frac{\kappa}{\sqrt{-g}}
\varepsilon^{mlnpq} {\cal F}_{ln} {\cal F}_{pq}
&=&
0,
\label{eom_a}
\end{eqnarray}
\end{subequations}

\noindent
where $T_{mn}$ is the energy-momentum tensor,
\begin{equation}
T_{mn}=\frac{1}{e^2}\Big({\cal F}_{mk}{\cal F}_{nl}g^{kl}
-\frac{1}{4}g_{mn}{\cal F}_{kl}{\cal F}^{kl}\Big).
\end{equation}
RN-AdS$_5$ background with 
AdS radius $l$ is a solution of the equations of motion (\ref{eom_g}) and
(\ref{eom_a}) even in the presence of the CS term:  
\begin{eqnarray}
\label{rnads}
\begin{array}{rcl}
(\dd s)^2
&=&
\displaystyle \frac{r^2}{l^2} \Big( -f(r)(\dd t)^2 +
(\dd\vec{x})^2\Big) + \frac{l^2}{r^2f(r)}(\dd r)^2,
\\
{\cal A}_t(r)
&=& \displaystyle - \frac{Q}{r^2} + \mu,
\end{array}
\quad
\end{eqnarray}
with
\begin{eqnarray*}
f(r) 
&=&
 1-\frac{ml^2}{r^4}+\frac{q^2l^2}{r^6}
\\
&=& 
\frac{1}{r^6}(r^2-r_0^2)(r^2-r_+^2)(r^2-r_-^2), 
\end{eqnarray*}
and 
$$
q=4\sqrt{\frac{\pi G_5}{3e^2}}Q.
$$
The parameters $m$ and $q$ correspond to the mass and charge of
the AdS space, respectively.
The asymptotic value of gauge field ${\cal A}_t(r\to\infty)=\mu$ 
may be interpreted as the chemical potential in the dual field theory. 
The explicit forms of $r_0(=-r_+^2-r_-^2)$ and $r_\pm$ are given by
\renewcommand{\arraystretch}{2.4}
\begin{eqnarray*}
\begin{array}{rcl}
r_0^2
&=&
\displaystyle \left( \frac{m}{3q^2} \Bigg( 1+2\cos \bigg(
               \frac{\theta}{3} +\frac{2}{3}\pi \bigg) \Bigg)
              \right)^{-1} , \label{r0}
\\
r_+^2
&=&
\displaystyle \left( \frac{m}{3q^2} \Bigg(
               1+2\cos\bigg(\frac{\theta}{3}+\frac{4}{3}\pi\bigg) \Bigg)
              \right)^{-1}, \label{r+}
\\
r^2_-
&=&
\displaystyle  \left( \frac{m}{3q^2}
\Bigg( 1+2\cos\bigg(\frac{\theta}{3}\bigg) \Bigg) \right)^{-1}, \label{r-}
\end{array}
\end{eqnarray*}
\renewcommand{\arraystretch}{1.7}
with
$$
\theta
=
\arctan \Bigg( \frac{3\sqrt{3}q^2\sqrt{\displaystyle
4m^3l^2-27q^4}}{2m^3l^2-27q^4} \Bigg),
$$
where
$r_+$ and $r_-$ represent
locations of the outer and inner horizons, respectively.

The Hawking temperature of RN-AdS$_5$ background 
which may correspond to the temperature of the dual field theory 
is given as
\begin{equation}
\label{temp}
T   =      \frac{r_+^2f'(r_+)}{4\pi l^2}
=      \frac{r_+}{\pi l^2}\bigg(1-\frac{1}{2}\frac{q^2l^2}{ r_+^6}\bigg)
\equiv \frac{1}{2\pi b}\Big(1-\frac{a}{2}\Big),
\end{equation}
where $a$ and $b$ are defined as
$$
a \equiv \frac{q^2l^2}{r_+^6}
\qquad  b \equiv \frac{l^2}{2r_+} \ .
$$
The value of $a$ can be taken in $0 \le a \le 2$. 
The entropy density $s$, the energy density $\epsilon$, the chemical
potential  and the physical 
charge density $\rho$ in the dual field theory are given by 
$$
s=\frac{r_+^3}{4G_5l^3}, \quad 
\epsilon=\frac{3m}{16G_5l^3}, \quad 
\mu=\frac{Q}{r_+^2}, \quad 
\rho=\frac{2Q}{e^2l^3}, 
$$
respectively. 
This kind of charged AdS background has been used in various contexts in 
the AdS/CFT correspondence~\cite{RN_AdS5}.    

It might be convenient to introduce new dimensionless coordinate
$u\equiv r_+^2/r^2$ which is normalized by the outer horizon.
In this coordinate, the horizon and the boundary are located at
$u=1$ and $u=0$, respectively.
The background (\ref{rnads}) can be rewritten as
\begin{eqnarray}
(\dd s)^2 
&=&
\frac{l^2}{4b^2u}\Big(-f(u)(\dd t)^2+(\dd\vec{x})^2\Big)
\nonumber 
\\
&&
+\frac{l^2}{4u^2f(u)}(\dd u)^2, 
\label{rnads_u}
\\
{\cal A}_t(u)
&=&
\mu(1-u),
\nonumber 
\end{eqnarray}
with 
$$
f(u)=(1-u)(1+u-au^2).
$$

\subsection{Zero temperature and AdS$_{\bm 2}$} 

We now consider the extremal limit i.e.\ 
$$
a=2. 
$$
Through the definition (\ref{temp}), 
we could access to the zero temperature system for the dual field 
theory. 
Hereafter we focus on this extremal zero temperature case. 
In this system, the only physical parameter in the bulk solution is the 
chemical potential $\mu$. 

Near the horizon $u=1$ in the extremal RN-AdS$_5$, 
the AdS$_2$ structure emerges~\cite{AdS2_NFL}. 
Near the horizon, 
we introduce dimensionless
coordinates $(\tau, \zeta)$  as
\begin{equation}
1-u=\frac{\epsilon}{\zeta},
\qquad 
t=\frac{\alpha}{\mu}\frac{\tau}{\epsilon}, 
\label{irads2}
\end{equation}
where we have introduced the dimensionless 
constant $\alpha=el/(4\sqrt{6\pi G_5})$ to make some expression simpler.
Taking the scaling limit $\epsilon\to 0$ with finite $(\tau, \zeta)$, 
the background (\ref{rnads_u}) becomes
\begin{equation}
\begin{array}{rcl}
(\dd s)^2
&=&
\displaystyle
\frac{l^2}{12\zeta^2}
\Big(-(\dd\tau)^2+(\dd\zeta)^2\Big)+\frac{l^2}{4b^2}(\dd\vec{x})^2,  
\\
A_\tau(\zeta)
&=&
\displaystyle
\frac{\alpha}{\zeta},
\end{array}
\label{ads2}
\end{equation}
which gives AdS$_2\times$R$^3$.   
We refer $\zeta\to\infty$ as the Poincar\'e horizon, while 
$\zeta\to 0$ as the AdS$_2$ boundary where the IR CFT could be defined.   

\section{Perturbations of RN-AdS$_{\bm 5}$ background} 

Now we consider perturbations on the extremal RN-AdS$_5$ background,
$$
g_{mn} = g^{(0)}_{mn}+h_{mn}
\quad {\rm and} \quad
{\cal A}_m = A_m^{(0)}+A_m,
$$
where $(g^{(0)}_{mn}(u),  A^{(0)}_m(u))$
and $(h_{mn}(u, x^\mu),   A_m(u, x^\mu))$ denote
the extremal RN-AdS$_5$ background (\ref{rnads_u}) and 
the perturbations, respectively.
We choose the following gauge conditions,
$$
\label{GC}
h_{um}=0 \quad \textrm{and} \quad A_u=0,
$$
and use Fourier expansion in which the momentum lies on the $z$-direction,
\renewcommand{\arraystretch}{2.4}
$$
\begin{array}{rcl}
h_{\mu\nu}(t, z, u)
&=&
\displaystyle \!\int\!\frac{\dd^2k}{(2\pi)^2} \
\mbox{e}^{-i\omega t+ikz}h_{\mu\nu}(\omega, k, u),
\\
A_\mu(t, z, u)
&=&
\displaystyle \!\int\!\frac{\dd^2k}{(2\pi)^2} \
\mbox{e}^{-i\omega t+ikz} A_\mu(\omega, k, u),
\end{array}
$$
where $\mu$ and $\nu$ run through 4D spacetime
except for the radial direction.
At the boundary, these fluctuations 
$(h_{\mu\nu}(u, x^\mu),  A_\mu(u, x^\mu))$  
may couple to the operators $(T^{\mu\nu}(x^\mu),  J^\mu(x^\mu))$ in 
the boundary field theory, where $T^{\mu\nu}(x^\mu)$ and $J^\mu(x^\mu)$ are 
the energy momentum tensor and
$U(1)$ charge current, respectively.  
By using Gubser-Klebanov-Polyakov-Witten (GKP-W) 
relation~\cite{AdS_CFT_Conjecture2}, 
one can calculate correlation functions through evaluating the bulk 
on-shell action. 
As we saw, the background solution does not depend on the existence 
of the CS term. 
On the other hand, fluctuations are affected by the CS term.  
The effect of the CS term to the dual field theory is our main topic. 

The perturbations can be categorized  to the three types
i.e.\ scalar, vector and tensor types by using the spin under the $O(2)$
rotation in the $(x,y)$-plane~\cite{pss}.
It is easy to show that 
the CS term contributes only to the vector type perturbation
whose non-zero values are listed below:
$$
h_{xt}, \quad h_{yt}, \quad
h_{xz}, \quad h_{yz}, \quad 
\mbox{and} \quad A_{x},  \quad A_{y}.
$$
It might be convenient to work with the variables 
$h^x_t(u)=g^{(0)xx}h_{xt}(u)$, $h^x_z(u)=g^{(0)xx}h_{xz}(u)$, 
$h^y_t(u)=g^{(0)yy}h_{yt}(u)$ and $h^y_z(u)=g^{(0)yy}h_{yz}(u)$. 

\subsection{Decoupling of the equation of motion}

The equations of motion (\ref{eom_g})
and (\ref{eom_a}) for the perturbation fields are given by 
\begin{subequations}
\begin{eqnarray}
0
&=&
{h^{x(y)}_t}''  - \frac{1}{u}{h^{x(y)}_t}' 
\nonumber 
\\
&&- \frac{9}{uf(u)}
\Big(\bm{\omega k} h^{x(y)}_z + \bm{k}^2h^{x(y)}_t \Big) - 6uB_{x(y)}', 
\qquad 
\label{eq_motion_v_001x} 
\\            
0
&=&
\bm{k}f(u){h^{x(y)}_z}' + \bm{\omega}{h^{x(y)}_t}'- 6\bm{\omega}u B_{x(y)},
\label{eq_motion_v_002x} 
\\             
0
&=&
{h^{x(y)}_z}'' + \frac{(u^{-1}f(u))'}{u^{-1}f(u)}{h^{x(y)}_z}'
\nonumber 
\\
&&
+ \frac{9}{uf^2(u)}
\Big(\bm{\omega}^2h^{x(y)}_z 
+\bm{\omega k}h^{x(y)}_t \Big), 
\label{eq_motion_v_003x} 
\end{eqnarray}
\end{subequations}
and
\begin{eqnarray}
0
&=&
B_{x(y)}'' + \frac{f'(u)}{f(u)}B_{x(y)}' + \frac{9}{uf^2(u)}
\Big(\bm{\omega}^2 -\bm{k}^2f(u) \Big) B_{x(y)} 
\nonumber 
\\
&&
- \frac{1}{f(u)}{h^{x(y)}_t}'
- (+)\tilde{\kappa} \frac{i \bm{k}}{f(u)}B_{y(x)},
\label{eq_motion_v_004x} 
\end{eqnarray}
with
\begin{eqnarray*}
&&f(u)=(1-u)^2(1+2u),
\\
&&
\tilde{\kappa} \equiv \frac{(12e)^2\alpha}{l}\kappa,
\quad
B_{x(y)} \equiv \frac{A_{x(y)}}{\mu}=\frac{b}{3\alpha}A_{x(y)}, 
\end{eqnarray*}
where the prime implies the derivative with respect to $u$. 
We have normalized  the frequency and momentum by the chemical potential 
$\bm{\omega}\equiv\alpha\omega/\mu$ 
and $\bm{k}\equiv\alpha k/\mu$, respectively. 
In the presence of the CS coupling $\tilde{\kappa}$, 
the $x$- and $y$-components of gauge fields are coupled. 
As we explain below, there are four independent variables.
Eq. (\ref{eq_motion_v_003x}) can be derived from
(\ref{eq_motion_v_001x}) and (\ref{eq_motion_v_002x}).
${h^{x(y)}_z}'(u)$ could be expressed in terms of ${h^{x(y)}_t}'(u)$ 
and $B_{x(y)}(u)$
through (\ref{eq_motion_v_002x}).
From (\ref{eq_motion_v_001x}) and (\ref{eq_motion_v_002x})
we can obtain a second order differential equation for ${h^{x(y)}_t}'(u)$ with
$B_{x(y)}(u)$.
Together with (\ref{eq_motion_v_004x}), 
we treat ${h^{x(y)}_t}'(u)$ and $B_{x(y)}(u)$ 
as the four independent variables.
  
In order to solve the coupled equations of motion,
it might be convenient to introduce master variables~\cite{ki}. 
By using master variables given in~\cite{mstt}, 
which correspond to the helicity bases on the $(x, y)$-plane~\cite{ty}, 
the equations of motion are decoupled and
organized as four ordinary differential equations. 
In the matrix form, the master equations are given by
\begin{equation}
0
=
\widetilde{\Theta}''
+\frac{(u^2f(u))'}{u^2f(u)}\widetilde{\Theta}'
+\widetilde{\Omega}(u)\widetilde{\Theta},
\label{eom_m}
\end{equation}
with potential 
\begin{eqnarray}
\widetilde{\Omega}(u)
&=&
\frac{9}{uf^2(u)}{\bm\omega}^2
+\frac{1}{2uf(u)}
\Big(
({\cal D}_{\tilde{\kappa}}\big(\bm{k})-6\big)u-18\bm{k}^2
\Big). 
\nonumber 
\\ 
\label{OMEGA}
\end{eqnarray}
The diagonal matrix ${\cal D}_{\tilde{\kappa}}(\bm{k})$ is given as
\begin{eqnarray}
&&{\cal D}_{\tilde{\kappa}}(\bm{k})
\nonumber 
\\
&&=\mbox{diag}
\Big(
-D_-+\tilde{\kappa}\bm{k}, \ D_-+\tilde{\kappa}\bm{k}, \
-D_+-\tilde{\kappa}\bm{k}, \ D_+-\tilde{\kappa}\bm{k}
\Big),
\nonumber 
\\
\end{eqnarray}
with constants $D_\pm$ defined through, 
\begin{eqnarray*}
C_\pm
&\equiv&
3\pm 3\sqrt{1+6\bm{k}^2},
\\
C_0
&\equiv&C_+-C_-, 
\\
D_{\pm}
&\equiv&
\sqrt{(C_0 \pm\tilde{\kappa}\bm{k})^2 \pm 4\tilde{\kappa}\bm{k}{C_-}}
\\
&=&
\sqrt{\tilde{\kappa}^2{\bm{k}}^2
+216{\bm k}^2\pm12\tilde{\kappa}{\bm k}+36}.  
\end{eqnarray*}
Here the master variables $\widetilde{\Theta}_a(u)$ are related with
the original variables via, 
\begin{equation}
\widetilde{\Theta}_a
=(\Lambda^{-1}\Theta)_a,
\label{inversion}
\end{equation}
where $\Theta_a\equiv (\Theta_{x+}, \Theta_{x-}, 
\Theta_{y+}, \Theta_{y-})^{\rm T}
\!\equiv(\Theta_1, \Theta_2, \Theta_3, \Theta_4)^{\rm T}$ 
are given as
\begin{equation}
\Theta_{x(y)\pm}
=
\frac{1}{u}{h^{x(y)}_t}'-\Big(6-\frac{C_\pm}{u}\Big)B_{x(y)}. 
\label{mixing}
\end{equation}
The constant matrix $\Lambda^{-1}$ is expressed as
\begin{widetext}
$$
\Lambda^{-1}
=
\left(
\begin{array}{cccc}
\displaystyle 
\ -2iC_-\tilde{\kappa}\bm{k} \
&
\displaystyle
\ -i(C_0^2-C_0D_--6\tilde{\kappa}\bm{k}) \
&
\displaystyle
\ -2C_-\tilde{\kappa}\bm{k} \
&
\displaystyle
\ -(C_0^2-C_0D_--6\tilde{\kappa}\bm{k}) \
\\
\displaystyle
\ 2iC_-\tilde{\kappa}\bm{k} \
&
\displaystyle
\ i(C_0^2+C_0D_--6\tilde{\kappa}\bm{k}) \
&
\displaystyle
\ 2C_-\tilde{\kappa}\bm{k} \
&
\displaystyle
\ C_0^2+C_0D_--6\tilde{\kappa}\bm{k} \
\\
\displaystyle
\ -2iC_-\tilde{\kappa}{\bm{k}} \ 
&
\displaystyle
\ i(C_0^2-C_0D_++6\tilde{\kappa}\bm{k}) \
&
\displaystyle
\ 2C_-\tilde{\kappa}{\bm{k}} \ 
&
\displaystyle
\ -(C_0^2-C_0D_++6\tilde{\kappa}\bm{k}) \ 
\\
\displaystyle
\ 2iC_-\tilde{\kappa}\bm{k} \ 
&
\displaystyle
\ -i(C_0^2+C_0D_++6\tilde{\kappa}\bm{k}) \ 
&
\displaystyle
\ -2C_-\tilde{\kappa}\bm{k} \ 
&
\displaystyle
\ C_0^2+C_0D_++6\tilde{\kappa}\bm{k}
\end{array}
\right). 
$$
\end{widetext}

We now apply the ``hydrodynamic'' analysis~\cite{ss, ss2} with small frequencies 
for the equations of motion.   
Compared with the finite temperature case~\cite{mstt}, 
in the extremal zero temperature case,  
the small $\bm{\omega}$ expansion does not work at the horizon which gives 
an irregular singularity. 
A sensible way has been developed in~\cite{AdS2_NFL} (see also \cite{gsw}).  
In~\cite{ejl}, Einstein-Maxwell theory on RN-AdS$_4$ background 
has been considered.   
In the low frequencies, 
we divide the holographic space labeled by $u$ into two parts so-called
``inner'' and ``outer'' regions and solve the equations of motion
independently.  
Then we match these solutions in the intermediate overlapping region 
to obtain full solutions.  

\subsection{Inner and outer regions, and matching}  

We first consider the equations of motion in the near horizon region.
By using the coordinate redefinitions (\ref{irads2}),  
the equations of motion (\ref{eom_m}) can be reduced to 
\begin{eqnarray}
0
&=&
\widetilde{\Theta}''_a
+\left(
{\bm\omega}_\tau^2-\frac{1}{\zeta^2}
\Big(1+3{\bm k}^2-\frac{1}{6}{\cal D}_{\tilde{\kappa}a}(\bm{k})\Big)
\right)\! 
\widetilde{\Theta}_a
+{\cal O}(\epsilon), 
\nonumber 
\\
\label{eom_ads2}
\end{eqnarray}
where we have defined the frequency in AdS$_2$ spacetime as 
\begin{equation}
{\bm\omega}_{\tau}\equiv\frac{\bm\omega}{\epsilon}, 
\label{omega}
\end{equation}
due to 
the scaling of the time coordinate in (\ref{irads2}).   
In order to keep ${\bm\omega}_\tau$ finite, 
the frequency ${\bm\omega}$ should be small i.e.\
${\bm\omega}<\epsilon$.  
We now define the inner region so that the equations of 
motion (\ref{eom_ads2}) could be well approximated: 
\begin{equation}
\mbox{Inner:} 
\quad 1-u<\epsilon\ll 1, \quad \mbox{with} \quad \bm{\omega}<\epsilon. 
\end{equation}

On the other hand, 
the outer region could be specified by introducing 
$\epsilon'(\gg \bm{\omega})$:   
\begin{equation}
\mbox{Outer:}\quad 
\bm{\omega}\ll \epsilon'<1-u,  
\end{equation}
in which the first term in the potential (\ref{OMEGA}) 
would be negligible compared to the other terms.   
Then the equations of motion (\ref{eom_m}) can be reduced in this 
outer region: 
\begin{eqnarray}
0
&=&
\widetilde{\Theta}_{a}''
+\frac{\Big(u^2(1-u)^2(1+2u)\Big)'}{u^2(1-u)^2(1+2u)}
\widetilde{\Theta}_{a}'
\nonumber 
\\
&&
+\frac{\big({\cal D}_{\tilde{\kappa}a}(\bm{k})-6\big)u
-18{\bm k}^2}{2u(1-u)^2(1+2u)}
\widetilde{\Theta}_{a} 
+{\cal O}({\bm\omega}).
\label{eom_o_0} 
\end{eqnarray}
In each regions, one can develop the solutions as a power series of 
$\bm{\omega}$.  

The overlapping region could be given through the double scaling limit,  
\begin{equation}
{\bm\omega}\ll\epsilon'<1-u<\epsilon\ll 1. 
\label{matching_region}
\end{equation}
One can match two solutions in the inner/outer in this overlapping region 
and may obtain full solutions.  

We now proceed to solve the equations of motion in each regions and 
to perform their matching. 
In the inner region, the leading equations in (\ref{eom_ads2}) are equivalent to 
that for massive scalar fields in AdS$_2$ spacetime. 
The effective AdS$_2$ masses are
\begin{equation}
l^2m^2_a
=1+3{\bm k}^2-\frac{1}{6}{\cal D}_{\tilde{\kappa}a}(\bm{k}). 
\end{equation}
This coincides with that discussed in~\cite{EM_CS_Nakamura}.  
In AdS spacetime, it is known that the mass is bounded below. 
This so-called  Breitenlohner-Freedman (BF) bound provides 
the critical value of the CS coupling~\cite{EM_CS_Nakamura}. 
We will discuss this instability issue later. 

Using the GKP-W relation~\cite{AdS_CFT_Conjecture2}, 
the conformal dimensions $\delta$ of operators which couple to the 
sources $\widetilde{\Theta}_{a}(u)$ are given by 
\begin{equation}
\delta_a=\frac{1}{2}+\Delta_{\tilde{\kappa}}^a(\bm{k}), 
\end{equation}
with 
\begin{eqnarray}
\Delta_{\tilde{\kappa}}^a(\bm{k}) 
&\equiv&
\sqrt{\left(\frac{1}{2}\right)^2+l^2m_a^2}
\nonumber 
\\
&=&
\frac{1}{2}\sqrt{5+12\bm{k}^2
-\frac{2}{3}{\cal D}_{\tilde{\kappa}a}(\bm{k})}. 
\label{Delta}
\end{eqnarray}
We can solve the leading equations of motion  in (\ref{eom_ads2})
analytically 
\begin{eqnarray}
\widetilde{\Theta}_{{\rm I} a}^{(0)}(\zeta)
&=&
A_a\sqrt{\zeta}\ J_{\Delta_{\tilde{\kappa}}^a(\bm{k})}
(\bm{\omega}_\tau\zeta)
+B_a\sqrt{\zeta}\ N_{\Delta_{\tilde{\kappa}}^a(\bm{k})}
(\bm{\omega}_\tau\zeta),
\nonumber 
\\
\end{eqnarray}
where $J_n(x)$ and $N_n(x)$ are Bessel functions of the first
and second kinds, respectively.
One of the integration constants $A_a$ and $B_a$ can be fixed
by imposing the in-coming wave condition at the horizon
$\zeta\to\infty$ where is deep inner region. 
Then we obtain  
\begin{equation}
\widetilde{\Theta}_{{\rm I}a}^{(0)}(\zeta)
=C_a\sqrt{\zeta}
\Big(
J_{\Delta_{\tilde{\kappa}}^a(\bm{k})}(\bm{\omega}_\tau\zeta)
+iN_{\Delta_{\tilde{\kappa}}^a(\bm{k})}(\bm{\omega}_\tau\zeta)
\Big).
\label{solution_ir}
\end{equation}
It should be noted that in the matching region (\ref{matching_region}) 
the condition $\bm{\omega}_\tau\zeta\ll1$ would be satisfied.   
Therefore in the matching region, we obtain the following asymptotic 
form of (\ref{solution_ir}),  
\begin{eqnarray}
\widetilde{\Theta}_{{\rm I}a}^{(0)}(\zeta)
\sim 
D_a
\bigg\{
&&(1-u)^{-\frac{1}{2}+\Delta_{\tilde{\kappa}}^a(\bm{k})}
\nonumber 
\\
&&
+
{\cal G}_{\bm{k} \tilde{\kappa}a}(\bm{\omega})
(1-u)^{-\frac{1}{2}-\Delta_{\tilde{\kappa}}^a(\bm{k})}
\bigg\}, \qquad 
\label{eom_i_1}
\end{eqnarray}
where we have used the relation given by (\ref{irads2}) and (\ref{omega}) 
\begin{equation}
\bm{\omega}_\tau\zeta=\frac{\bm{\omega}}{1-u}, 
\label{iruv}
\end{equation}
to get expressions in the original $u$-coordinate 
and also introduced the overall normalization constant
$D_a$.
Here ${\cal G}_{{\bm k}\tilde{\kappa}a}(\bm{\omega})$ can be 
estimated as
\begin{eqnarray}
&&
{\cal G}_{{\bm k}\tilde{\kappa} a}(\bm{\omega})
\nonumber 
\\
&&
\hspace*{5mm}
=
-{\rm e}^{-i\pi\Delta_{\tilde{\kappa}}^a(\bm{k})}
\frac{\Gamma\Big(1-\Delta_{\tilde{\kappa}}^a(\bm{k})\Big)}
{\Gamma\Big(1+\Delta_{\tilde{\kappa}}^a(\bm{k})\Big)}
\Big(\frac{\bm{\omega}}{2}\Big)^{2\Delta_{\tilde{\kappa}}^a(\bm{k})},
\quad 
\end{eqnarray}
which may be related with the retarded correlation functions of 
the IR CFT~\cite{AdS2_NFL},
\begin{equation}
G_{\rm IR}(\bm{\omega}, \bm{k})_{\tilde{\kappa}}
\sim
\Delta_{\tilde{\kappa}}^a(\bm{k})
{\cal G}_{{\bm k}\tilde{\kappa}a}(\bm{\omega}).
\end{equation}
It is easy to see that through the relation (\ref{iruv}) 
the matching region (also inner region) covers the outer region.  
Hence we could expect to have an appropriate  matching between the 
solutions.   

In the outer region, 
we need to solve the leading equation of motion (\ref{eom_o_0}). 
From the right hand side of the inequality in (\ref{matching_region}), 
the near horizon region around $u=1$ can be understood as the 
matching region from the outer region. 
We can factorize the regular singularity around the horizon as   
\begin{equation}
\widetilde{\Theta}_{{\rm O}a}^{(0)}(u)
=(1-u)^{\nu_a} F_a(u), 
\end{equation}
with
\begin{equation}
\nu_a=\nu_{a\pm}=-\frac{1}{2}\pm\Delta_{\tilde{\kappa}}^a(\bm{k}). 
\label{exponent}
\end{equation}
The function $F_a(u)$ follows the Heun equation~\footnote{
%%% 
The Heun equation is generally given as
\\
\\
$\displaystyle
0
=
{\cal W}''
+\Big\{\frac{\gamma}{z}+\frac{\delta}{z-1}+\frac{\epsilon}{z-a}
\Big\}{\cal W}'
+\frac{\alpha\beta z-q}{z(z-1)(z-a)}{\cal W},
$
\\
\\
with 
$
\epsilon=\alpha+\beta-\gamma-\delta+1.
$
Under the assignment of the parameters;
$$
\alpha=\nu_a, \quad 
\beta=\nu_a+4, \quad 
\gamma=2, \quad
\delta=2(1+\nu_a), 
$$
$$
a=-\frac{1}{2}, \quad
q=-\nu_a-\frac{9}{2}{\bm k}^2,
$$
we could have the form (\ref{eom_o_1}).
},   
\begin{eqnarray}
0
&=&
F_a''(u)
+\bigg\{\frac{2}{u}+\frac{2(1+\nu_a)}{u-1}
+\frac{1}{\displaystyle u+1/2}\bigg\}F_a'(u)
\nonumber 
\\
&&
+\frac{\displaystyle\nu_a(\nu_a+4)u+\nu_a+\frac{9}{2}{\bm k}^2}
{\displaystyle u(u-1)(u+1/2)}F_a(u). 
\label{eom_o_1}
\end{eqnarray}
The Heun functions which are solutions of the Heun equation are
not among the usual special functions.
In order to solve the Heun equation in the entire region,
one needs to use some numerical methods.

The exponents (\ref{exponent}) are the same as those in 
(\ref{eom_i_1}). 
Therefore using the following functions 
\begin{equation}
\eta^{(0)\pm}_a(u)
=(1-u)^{-\frac{1}{2}\pm\Delta_{\tilde{\kappa}}^a(\bm{k})}
\Big(1+{\cal O}(1-u)\Big), 
\label{basis_horizon}
\end{equation}
as two linear independent outer solutions in the matching region,  
the matching with the inner solution (\ref{eom_i_1}) could be
carried out by fixing coefficients for linear combinations of the
basis functions (\ref{basis_horizon}). 
Hence we have 
\begin{equation}
\widetilde{\Theta}^{(0)}_{{\rm O}a}(u)
=
D_a\Big\{\eta^{(0)+}_a(u)
+{\cal G}_{{\bm k}\tilde{\kappa}a}(\bm{\omega})\eta^{(0)-}_a(u)
\Big\}.  
\label{om}
\end{equation}

Near the boundary $u=0$,  the indicial exponents of (\ref{eom_o_0}) (and
(\ref{eom_o_1})) are $(0, -1)$.
We take the following Frobenius series solutions as two independent
basis,
\begin{equation}
\begin{array}{rcl}
\xi^{({\rm I})}(u)
&=&
1+{\cal O}(u), 
\\
\xi^{({\rm II})}(u)
&=&
\displaystyle
\frac{1}{u}
\Big(1+{\cal O}(u^2)\Big)
+9{\bm k}^2\xi^{({\rm I})}(u)\log u, 
\qquad 
\end{array}
\label{asymptotic}
\end{equation}
which correspond to the normalizable and nonnormalizable modes, 
respectively~\cite{bkl}. 
We can develop the solutions (\ref{om}) to the boundary and may express
those by 
the asymptotic expansion of the two basis 
solutions (\ref{asymptotic}), 
\begin{eqnarray}
&&
\hspace*{-1mm}
\widetilde{\Theta}^{(0)}_{{\rm O}a}(u)
\nonumber 
\\
&&
=
D_a
\Bigg\{
\Big({\cal A}_{\tilde{\kappa}a}^+(\bm{k})\xi^{({\rm I})}(u)
+{\cal B}_{\tilde{\kappa}a}^+(\bm{k})\xi^{({\rm II})}(u)\Big)
\nonumber 
\\
&&
\hspace*{11mm}
+
{\cal G}_{\bm{k}\tilde{\kappa}a}(\bm{\omega})
\Big({\cal A}_{\tilde{\kappa}a}^-(\bm{k})\xi^{({\rm I})}(u)
+{\cal B}_{\tilde{\kappa}a}^-(\bm{k})\xi^{({\rm II})}(u)\Big)
\Bigg\}. 
\nonumber 
\\
&&
\label{outer_final}
\end{eqnarray}
Although there are no analytic expressions of the connection coefficients
${\cal A}_{\tilde{\kappa}a}^\pm(\bm{k})$ and 
${\cal B}_{\tilde{\kappa}a}^\pm(\bm{k})$,  
one could discuss the low frequency behavior of the correlation
functions~\cite{AdS2_NFL}.   
Following this direction, we proceed to obtain the correlation 
functions ``formally''. 
One can also consider higher order corrections in 
the small $\bm{\omega}$-expansions~\cite{AdS2_NFL}.  
The connection coefficients could be determined perturbatively, 
\begin{eqnarray}
&&
\hspace*{-3mm}
{\cal A}^\pm_{\tilde{\kappa}a}(\bm{k})
\nonumber 
\\
&&\longrightarrow
{\cal A}_{\tilde{\kappa}a}^\pm(\bm{\omega}, \bm{k})
={\cal A}^{\pm(0)}_{\tilde{\kappa}a}(\bm{k})
+ \bm{\omega}{\cal A}^{\pm(1)}_{\tilde{\kappa}a}(\bm{k})
+{\cal O}({\bm{\omega}}^2), 
\nonumber 
\\
&&
\hspace*{-3mm}
{\cal B}^\pm_{\tilde{\kappa}a}(\bm{k})
\nonumber 
\\
&&\longrightarrow
{\cal B}^\pm_{\tilde{\kappa}a}(\bm{\omega}, \bm{k})
={\cal B}^{\pm(0)}_{\tilde{\kappa}a}(\bm{k})
+ \bm{\omega}{\cal B}^{\pm(1)}_{\tilde{\kappa}a}(\bm{k})
+{\cal O}({\bm{\omega}}^2). 
\nonumber 
\\
\label{corrections}
\end{eqnarray}
However, in this paper, it is enough to consider only the leading contributions.   

The remaining thing is to fix the overall constants $D_a$ in the
solutions.  
These can be determined in terms of the field values
at the boundary. 
In order for that, we use relations derived by (\ref{eq_motion_v_001x}) and
(\ref{mixing}),
\begin{eqnarray*}
&&\displaystyle u^2 \Theta_{x(y)\pm}'-uC_\pm B_{x(y)}' \Big|_{u=0}
\\
&=&
9
\Big(
\bm{k}^2 \big(h^{x(y)}_t\big)^{(0)}
+\bm{\omega k}\big(h^{x(y)}_z\big)^{(0)}
\Big)
-C_\pm\big(B_{x(y)}\big)^{(0)}, 
\end{eqnarray*}
where $\big(h^{x(y)}_t\big)^{(0)}$, $\big(h^{x(y)}_z\big)^{(0)}$ 
and $\big(B_{x(y)}\big)^{(0)}$
stand for their constant values at the boundary.
Hence we could obtain the overall constants as: 
%
%%% new arraystretch
\renewcommand{\arraystretch}{3.6}
%%%
\begin{widetext}
\begin{equation}
\begin{array}{rcl}
D_{1}
&=&
\displaystyle
\frac{9\widetilde{K}_-
\Big(\bm{k}^2\left(i(h^x_t)^{(0)}+(h^y_t)^{(0)}\right)
+\bm{\omega k}\left(i(h^x_z)^{(0)}+(h^y_z)^{(0)}\right)
\Big)
-C_-\widetilde{K}_+\Big(i(B_x)^{(0)}+(B_y)^{(0)}\Big)}
{
\Big({\cal B}_{\tilde{\kappa}1}^+({\bm k})
+{\cal G}_{\bm{k}\tilde{\kappa}1}(\bm{\omega})
{\cal B}_{\tilde{\kappa}1}^-(\bm{k})
\Big)},
\\
D_{2}
&=&
\displaystyle
\frac{-9K_-
\Big({\bm k^2}\left(i(h^x_t)^{(0)}+(h^y_t)^{(0)}\right)
+{\bm{\omega k}}\left(i(h^x_z)^{(0)}+(h^y_z)^{(0)}\right)
\Big)
+C_-K_+\Big(i(B_x)^{(0)}+(B_y)^{(0)}\Big)}
{\Big({\cal B}_{\tilde{\kappa}2}^+(\bm{k})
+{\cal G}_{\bm{k}\tilde{\kappa}2}(\bm{\omega})
{\cal B}_{\tilde{\kappa}2}^-(\bm{k})\Big)},
\\
D_{3}
&=&
\displaystyle
\frac{-9\widetilde{L}_-
\Big({\bm k^2}\left(i(h^x_t)^{(0)}-(h^y_t)^{(0)}\right)
+{\bm{\omega k}}\left(i(h^x_z)^{(0)}-(h^y_z)^{(0)}\right)\Big)
+C_-\widetilde{L}_+\Big(i(B_x)^{(0)}-(B_y)^{(0)}\Big)}
{\Big({\cal B}_{\tilde{\kappa}3}^+(\bm{k})
+{\cal G}_{\bm{k}\tilde{\kappa}3}(\bm{\omega})
{\cal B}_{\tilde{\kappa}3}^-(\bm{k})\Big)},
\\
D_{4}
&=&
\displaystyle
\frac{9L_-
\Big({\bm k^2}\left(i(h^x_t)^{(0)}-(h^y_t)^{(0)}\right)
+{\bm{\omega k}}\left(i(h^x_z)^{(0)}-(h^y_z)^{(0)}\right)\Big)
-C_-L_+\Big(i(B_x)^{(0)}-(B_y)^{(0)}\Big)}
{\Big({\cal B}_{\tilde{\kappa}4}^+(\bm{k})
+{\cal G}_{\bm{k}\tilde{\kappa}4}(\bm{\omega})
{\cal B}_{\tilde{\kappa}4}^-(\bm{k})\Big)},
\end{array}
\end{equation}
\end{widetext}
with 
%
%%% new arraystretch
\renewcommand{\arraystretch}{1.6}
%%%
\begin{equation}
\begin{array}{rcl}
K_{\pm}(\bm{k}, \tilde{\kappa})
&\equiv&
C_0^2(\bm{k})+D_-(\tilde{\kappa}, \bm{k})C_0(\bm{k})
\\
&&
+2\tilde{\kappa}\Big(C_{\pm}(\bm{k})-3\Big){\bm k}, 
\\
\widetilde{K}_{\pm}(\bm{k}, \tilde{\kappa})
&\equiv&
C_0^2(\bm{k})-D_-(\tilde{\kappa}, \bm{k})C_0(\bm{k})
\\
&&
+2\tilde{\kappa}\Big(C_{\pm}(\bm{k})-3\Big){\bm k}, 
\\
L_{\pm}(\bm{k}, \tilde{\kappa})
&\equiv&
C_0^2(\bm{k})+D_+(\tilde{\kappa}, \bm{k})C_0(\bm{k})
\\
&&
-2\tilde{\kappa}\Big(C_{\pm}(\bm{k})-3\Big){\bm k},
\\
\widetilde{L}_{\pm}(\bm{k}, \tilde{\kappa})
&\equiv&
C_0^2(\bm{k})-D_+(\tilde{\kappa}, \bm{k})C_0(\bm{k})
\\
&&
-2\tilde{\kappa}\Big(C_{\pm}(\bm{k})-3\Big){\bm k}.
\end{array}
\label{K}
\end{equation}
We have formally fixed the leading order solutions of the master variables 
$\widetilde{\Theta}^{(0)}_a(u)$.    

\section{Retarded correlation functions} 

In order to obtain the two-point retarded correlation function
via the GKP-W relation~\cite{AdS_CFT_Conjecture2},
we need the bilinear-part of the regularized on-shell action at the
boundary $u=0$.
Here we can prepare the following counter terms for the UV 
regularization~\cite{bk}:  
\begin{eqnarray}
S_{\rm ct} = S_{\rm ct}^{\rm gravity}+S_{\rm ct}^{\rm gauge},
\label{ct}
\end{eqnarray}
with 
\begin{subequations}
\begin{eqnarray*}
S_{\rm ct}^{\rm gravity}
&=&
\frac{1}{8\pi G_5}
\Bigg\{\! \int \! \dd^4x \sqrt{-g^{(4)}}
\Big(\ \frac{3}{l} + \frac{l}{4} R^{(4)}\Big)
\\
&&
\hspace*{10mm}
-\frac{l^3}{16}\log u \! \int \! \dd^4x \sqrt{-g^{(4)}}
\Big(R_{\mu\nu}^{(4)}R^{(4)\mu\nu}
\\
&&
\hspace*{47mm}
-\frac{1}{3}\big(R^{(4)}\big)^2\Big)
\Bigg\}, 
\\
S_{\rm ct}^{\rm gauge}
&=&
 \frac{l}{8e^2} \log u \! \int \! \dd^4x \sqrt{-g^{(4)}}
\ {\cal F}_{\mu\nu} {\cal F}^{\mu\nu},
\end{eqnarray*}
\end{subequations}

\noindent
where $R^{(4)}$ and $R^{(4)}_{\mu\nu}$ are the scalar curvature and
the Ricci tensor on the 4D boundary, respectively.

The bilinear parts of the perturbations in the bulk actions 
(\ref{eh}), (\ref{maxwell}) and (\ref{cs}) are reduced to the surface
terms by using the equations of motion.
In the present gauge, the extrinsic curvature in the Gibbons-Hawking
term (\ref{gh}) is given as 
$K=g^{(4)\mu\nu}\del_ug^{(4)}_{\mu\nu}/(2\sqrt{g_{uu}})$.  
Then in the low frequencies, 
the regularized boundary action is given by 
\begin{widetext}
\begin{eqnarray}
\label{surface}
S_{\rm on-shell}
&=&
\lim_{u \to 0}\Big(S + S_{\rm ct}\Big)\Big|_{\rm on-shell}
\nonumber
\\
&=& \lim_{u \to 0} 
\frac{l^3}{256\pi b^4 G_5}\!\int\!\frac{\dd^2k}{(2\pi)^2}
\Bigg\{
\frac{1}{u}\Big(h^x_t(-k, u){h^x_t}'(k, u)
-h^x_z(-k, u){h^x_z}'(k, u)\Big)
\nonumber
\\
&&
\hspace*{37mm}
+\frac{3}{2}\Big(3h^x_t(-k, u)h^x_t(k, u)
+h^x_z(-k, u)h^x_z(k, u)\Big)
\nonumber \\ 
&&
\hspace*{37mm}
-6B_x(-k, u){B_x}'(k, u)+6B_x(-k, u)h^x_t(k, u)
\nonumber 
\\
&&
\hspace*{37mm}
+\Big(\frac{9}{u}
+81\bm{k}^2\log u
\Big)
\nonumber
\\
&&
\hspace*{42mm}
\times
\Big({\bm k}^2h^x_t(-k, u)h^x_t(k, u)
+{\bm\omega}^2h^x_z(-k, u)h^x_z(k, u)
+2\bm{\omega k}h^x_t(-k, u)h^x_z(k, u)\Big)
\nonumber 
\\
&&
\hspace*{37mm}
+54\bm{k}^2
\log u B_x(-k, u)B_x(k, u)
+(x\to y) 
\Bigg\}.
\end{eqnarray}
\end{widetext}

Performing the matrix inversion (\ref{inversion}) and (\ref{mixing}) to 
rewrite the solutions in terms of 
the original variables, and following the prescription given in
Appendix,  
we could obtain two-point retarded correlation functions: 
\begin{subequations}
\begin{eqnarray}
G_{x \ x}(\bm{\omega}, \bm{k})_{\tilde{\kappa}}
&=&
-\frac{l}{12e^2b^2}
\bigg\{
54{\bm{k}}^2+\frac{3}{4\tilde{\kappa}C_0^2\bm{k}}
\widetilde{G}_{1+}(\bm{\omega}, \bm{k})_{\tilde{\kappa}}
\bigg\}, 
\nonumber 
\\
&&
\\
G_{x \ y}({\bm{\omega}}, \bm{k})_{\tilde{\kappa}}
&=&
\frac{l}{12e^2b^2}
\Big(
\frac{3i}{8\tilde{\kappa}C_0^2\bm{k}}
\Big)\widetilde{G}_{2+}(\bm{\omega}, \bm{k})_{\tilde{\kappa}}, 
\\
G_{xt \ xt}(\bm{\omega}, \bm{k})_{\tilde{\kappa}}
&=&
\frac{l^3}{128\pi b^4G_5}
\bigg\{
\frac{9}{2}
-81{\bm k}^4
\nonumber 
\\
&&
\hspace*{18mm}
+\frac{9\bm{k}}{8\tilde{\kappa}C_0^2}
\widetilde{G}_{1-}(\bm{\omega}, \bm{k})_{\tilde{\kappa}}
\bigg\},
\\
G_{xz \ xz}(\bm{\omega}, \bm{k})_{\tilde{\kappa}}
&=&
\frac{l^3}{128\pi b^4G_5}
\bigg\{
\frac{3}{2}
-81\bm{\omega}^2{\bm k}^2
\nonumber 
\\
&&
\hspace*{17mm}
+\frac{9\bm{\omega}^2}{8\tilde{\kappa}C_0^2\bm{k}}
\widetilde{G}_{1-}(\bm{\omega}, \bm{k})_{\tilde{\kappa}}
\bigg\},
\\
G_{xt \ xz}(\bm{\omega}, \bm{k})_{\tilde{\kappa}}
&=&
\frac{l^3}{256\pi b^4G_5}
\nonumber 
\\
&&
\hspace*{2mm}
\times\bigg\{
81\bm{\omega}\bm{k}^4
-\frac{9\bm{\omega}}{8\tilde{\kappa}C_0^2}
\widetilde{G}_{1-}(\bm{\omega}, \bm{k})_{\tilde{\kappa}}
\bigg\},
\\
G_{xt \ yt}(\bm{\omega}, \bm{k})_{\tilde{\kappa}}
&=&
-\frac{l^3}{256\pi b^4G_5}
\Big(\frac{9i\bm{k}}{8\tilde{\kappa}C_0^2}\Big)
\widetilde{G}_{2-}(\bm{\omega}, \bm{k})_{\tilde{\kappa}}, 
\\
G_{xt \ yz}(\bm{\omega}, \bm{k})_{\tilde{\kappa}}
&=&
\frac{l^3}{256\pi b^4G_5}
\Big(
\frac{9i\bm{\omega}}{8\tilde{\kappa}C_0^2}
\Big)\widetilde{G}_{2-}(\bm{\omega}, \bm{k})_{\tilde{\kappa}}, 
\\
G_{xz \ yz}(\bm{\omega}, \bm{k})_{\tilde{\kappa}}
&=&
-\frac{l^3}{256\pi b^4G_5}
\Big(
\frac{9i\bm{\omega}^2}{8\tilde{\kappa}C_0^2\bm{k}}
\Big)\widetilde{G}_{2-}(\bm{\omega}, \bm{k})_{\tilde{\kappa}}, 
\qquad
\end{eqnarray}
\begin{eqnarray}
G_{x \ xt}(\bm{\omega}, \bm{k})_{\tilde{\kappa}}
&=&
-\frac{l^2}{32e\sqrt{6\pi G_5}b^3}
\nonumber 
\\
&&
\hspace*{2mm}
\times
\bigg\{
6
+\frac{27\bm{k}}{4\tilde{\kappa}C_0^2C_-}
\widetilde{G}_{3-}(\bm{\omega}, \bm{k})_{\tilde{\kappa}}
\bigg\}, 
\\
G_{x \ xz}(\bm{\omega}, \bm{k})_{\tilde{\kappa}}
&=&
\frac{l^2}{32e\sqrt{6\pi G_5}b^3}
\Big(
\frac{27\bm{\omega}}{4\tilde{\kappa}C_0^2C_-}
\Big)
\widetilde{G}_{3-}(\bm{\omega}, \bm{k})_{\tilde{\kappa}}, 
\nonumber 
\\
&&
\\
G_{x \ yt}(\bm{\omega}, \bm{k})_{\tilde{\kappa}}
&=&
\frac{l^3}{32e\sqrt{6\pi G_5}b^3}
\Big(
\frac{27i\bm{k}}{4\tilde{\kappa}C_0^2C_-}
\Big)\widetilde{G}_{3+}(\bm{\omega}, \bm{k})_{\tilde{\kappa}}, 
\nonumber 
\\
&&
\\
G_{x \ yz}(\bm{\omega}, \bm{k})_{\tilde{\kappa}}
&=&
-\frac{l^2}{32e\sqrt{6\pi G_5}b^3}
\Big(
\frac{27i\bm{\omega}}{4\tilde{\kappa}C_0^2C_-}
\Big)\widetilde{G}_{3+}(\bm{\omega}, \bm{k})_{\tilde{\kappa}}, 
\nonumber 
\\
&&
\end{eqnarray}
\end{subequations}
with
{\small
\begin{eqnarray*}
\widetilde{G}_{1\pm}(\bm{\omega}, \bm{k})_{\tilde{\kappa}}
&=&
\!
\Big(\frac{\widetilde{K}_\pm K_\mp}{D_-}\Big)
G_1(\bm{\omega}, \bm{k})_{\tilde{\kappa}}
-\Big(\frac{\widetilde{K}_\mp K_\pm}{D_-}\Big)
G_2(\bm{\omega}, \bm{k})_{\tilde{\kappa}}
\nonumber 
\\
&&
-\Big(\frac{\widetilde{L}_\pm L_\mp}{D_+}\Big)
G_3(\bm{\omega}, \bm{k})_{\tilde{\kappa}}
+\Big(\frac{\widetilde{L}_\mp L_\pm}{D_+}\Big)
G_4(\bm{\omega}, \bm{k})_{\tilde{\kappa}}, 
\nonumber 
\\
\widetilde{G}_{2\pm}(\bm{\omega}, \bm{k})_{\tilde{\kappa}}
&=&
\!
\Big(\frac{\widetilde{K}_\pm K_\mp}{D_-}\Big)
G_1(\bm{\omega}, \bm{k})_{\tilde{\kappa}}
-\Big(\frac{\widetilde{K}_\mp K_\pm}{D_-}\Big)
G_2(\bm{\omega}, \bm{k})_{\tilde{\kappa}}
\nonumber 
\\
&&
+\Big(\frac{\widetilde{L}_\pm L_\mp}{D_+}\Big)
G_3(\bm{\omega}, \bm{k})_{\tilde{\kappa}}
-\Big(\frac{\widetilde{L}_\mp L_\pm}{D_+}\Big)
G_4(\bm{\omega}, \bm{k})_{\tilde{\kappa}}, 
\nonumber 
\\
\widetilde{G}_{3\pm}(\bm{\omega}, \bm{k})_{\tilde{\kappa}}
&=&
\Big(\frac{\widetilde{K}_-K_-}{D_-}\Big)
\Big(
G_1(\bm{\omega}, \bm{k})_{\tilde{\kappa}}
-G_2(\bm{\omega}, \bm{k})_{\tilde{\kappa}}
\Big)
\nonumber 
\\
&&
\pm\Big(\frac{\widetilde{L}_-L_-}{D_+}\Big)
\Big(
G_3(\bm{\omega}, \bm{k})_{\tilde{\kappa}}
-G_4(\bm{\omega}, \bm{k})_{\tilde{\kappa}}
\Big), 
\end{eqnarray*}
}
and  
\begin{eqnarray}
&&
\hspace*{-5mm}
G_a(\bm{\omega}, \bm{k})_{\tilde{\kappa}}
\nonumber 
\\
&=&
\frac{{\cal A}_{\tilde{\kappa}a}^+(\bm{k})
+{\cal G}_{\bm{k}\tilde{\kappa}a}(\bm{\omega})
{\cal A}_{\tilde{\kappa}a}^-(\bm{k})
}
{{\cal B}_{\tilde{\kappa}a}^+(\bm{k})
+{\cal G}_{\bm{k}\tilde{\kappa}a}(\bm{\omega})
{\cal B}_{\tilde{\kappa}a}^-(\bm{k})}
\nonumber 
\\
&\sim&
\frac{{\cal A}^+_{\tilde{\kappa}a}(\bm{k})}
{{\cal B}^+_{\tilde{\kappa}a}(\bm{k})}
\Bigg\{
1+
\left(\frac{{\cal A}^-_{\tilde{\kappa}a}(\bm{k})}
{{\cal A}^+_{\tilde{\kappa}a}(\bm{k})}
-\frac{{\cal B}^-_{\tilde{\kappa}a}(\bm{k})}
{{\cal B}^+_{\tilde{\kappa}a}(\bm{k})}
\right){\cal G}_{\bm{k}\tilde{\kappa}a}(\bm{\omega})
+\cdots
\Bigg\}, 
\nonumber 
\\
\label{Gi}
\end{eqnarray}
where $K_\pm$, $\widetilde{K}_\pm$ $L_\pm$ and $\widetilde{L}_\pm$ are 
given in (\ref{K}). 
In the last line in (\ref{Gi}), we did small ${\bm\omega}$
expansions assuming ${\cal B}_{\tilde{\kappa}a}^+ \ne 0$.    
We can observe non-analytical frequency behavior 
of the correlation functions which is determined by the IR CFT 
i.e.\ 
\begin{equation}
{\cal G}_{\bm{k}\tilde{\kappa}a}(\bm{\omega})
\propto{\bm{\omega}}^{2\Delta_{\tilde{\kappa}}^a(\bm{k})}.
\end{equation}
UV informations are given by the 
coefficients ${\cal A}^\pm_{\tilde{\kappa}a}(\bm{k})$ and 
${\cal B}^\pm_{\tilde{\kappa}a}(\bm{k})$;  
more generally ${\cal A}^\pm_{\tilde{\kappa}a}(\bm{\omega}, \bm{k})$ and 
${\cal B}^\pm_{\tilde{\kappa}a}(\bm{\omega}, \bm{k})$  in
(\ref{corrections})~\cite{AdS2_NFL}.       

\section{Discussion}

It has been discussed that the presence of the CS term can cause
the uniform solution without the CS term to be unstable against
non-uniform current ordering above the critical value $\tilde{\kappa}_c$ 
determined from the Breitenlohner-Freedman (BF) bound for
fluctuations \cite{EM_CS_Nakamura}. 
Indeed, the CS term gives rise 
to couplings between the $x$- and $y$- components of the
``effective'' vector potential with ``anomalous'' correlations. 
This indicates that dual $U(1)$ charge currents may feel anomalous 
interactions between not only $x$- or $y$- components 
but also $x$- and $y$-components, 
allowing us to expect possible instability against
non-uniform current ordering. 
If the CS term is turned off, such correlations between 
effective gauge fields disappear and their correlations become trivial. 
Although the CS coefficient cannot be used as a tuning parameter, 
we speculate that the corresponding CFT 
with $\tilde{\kappa}\approx\tilde{\kappa}_c$ differs from
that with $\tilde{\kappa} < \tilde{\kappa}_c$ in the respect 
that the CFT of $\tilde{\kappa} < \tilde{\kappa}_c$ may 
describe uniform current fluctuations
from the trivial vacuum state while that of 
$\tilde{\kappa}\approx\tilde{\kappa}_c$ would describe complex 
current fluctuations with anomalous correlations.
It is interesting to observe that the CS coefficient given by the
string theory is almost identical to the critical value
\cite{EM_CS_Nakamura}.

An essential question is the role of the CS term in critical
exponents of $\mathcal{G}_{\bm{k}\tilde{\kappa}a}(\bm{\omega})$, 
i.e.\ $\Delta_{\tilde{\kappa}}^a(\bm{k})$ given by (\ref{Delta}).   
In the absence of
the CS term we obtain 
$\Delta_{\tilde{\kappa}=0}^1(\bm{k})
=\Delta_{\tilde{\kappa}=0}^{3} (\bm{k})
>\Delta_{\tilde{\kappa}=0}^{2} (\bm{k}) 
= \Delta_{\tilde{\kappa}=0}^{4} (\bm{k})$ 
as shown in the right panel of Fig.1. 
This is quite natural because the conformal
dimensions of the $x$-current is expected to be the same as that of
the $y$-current. 
On the other hand, the CS term makes 
$\Delta_{\tilde{\kappa}=0}^{1} (\bm{k}) 
= \Delta_{\tilde{\kappa}=0}^{3} (\bm{k})$ 
split into 
$\Delta_{\tilde{\kappa}}^{1} (\bm{k}) 
< \Delta_{\tilde{\kappa}=0}^{1} (\bm{k}) 
=\Delta_{\tilde{\kappa}=0}^{3} (\bm{k}) 
< \Delta_{\tilde{\kappa}}^{3} (\bm{k})$. 
The same thing happens for 
$\Delta_{\tilde{\kappa}=0}^{2} (\bm{k}) 
= \Delta_{\tilde{\kappa}=0}^{4} (\bm{k})$. 
This becomes $\Delta_{\tilde{\kappa}}^{2} (\bm{k}) 
<\Delta_{\tilde{\kappa}=0}^{2} (\bm{k}) 
= \Delta_{\tilde{\kappa}=0}^{4} (\bm{k}) 
<\Delta_{\tilde{\kappa}}^{4} (\bm{k})$ 
by the CS term. 
These are shown in  the left panel of Fig.2.
As discussed before, the presence of the CS term is expected to cause
the instability to the non-uniform current ordering. 
This effect can be seen from the fact that the BF bound is translated into the
minimum value of $\Delta_{\tilde{\kappa}}^{2} (\bm{k}_{c})$, as shown
in the left panel of Fig.1. 
As a result, we find
$\Delta_{\tilde{\kappa}}^{2} (\bm{k}_{c}) 
< \Delta_{\tilde{\kappa}=0}^{2} (\bm{k}_{c}) 
=\Delta_{\tilde{\kappa}=0}^{4} (\bm{k}_{c}) 
< \Delta_{\tilde{\kappa}}^{4} (\bm{k}_{c}) 
<\Delta_{\tilde{\kappa}}^{1} (\bm{k}_{c}) 
< \Delta_{\tilde{\kappa}=0}^{1} (\bm{k}_{c}) 
=\Delta_{\tilde{\kappa}=0}^{3} (\bm{k}_{c}) 
< \Delta_{\tilde{\kappa}}^{3} (\bm{k}_{c})$ 
for critical fluctuations.
Novel critical exponents appear in the current-current correlation
function of critical fluctuations as a result of the interplay
between the CS term and the emergent locality.

\begin{figure}% [b]
\includegraphics[width=0.23\textwidth]{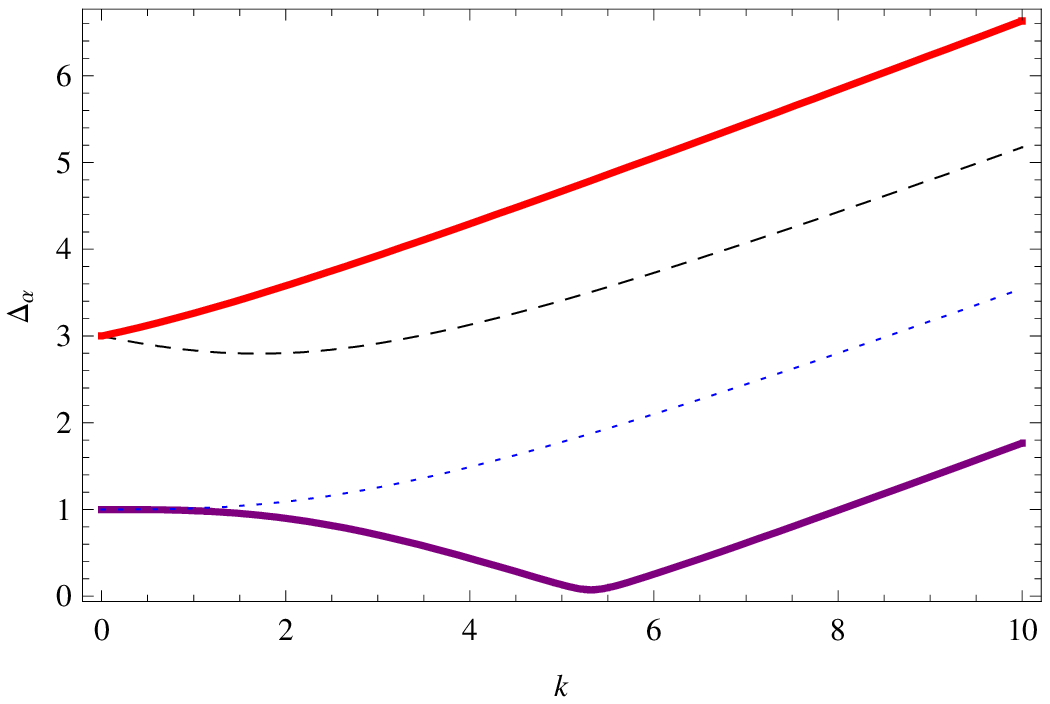}
\includegraphics[width=0.23\textwidth]{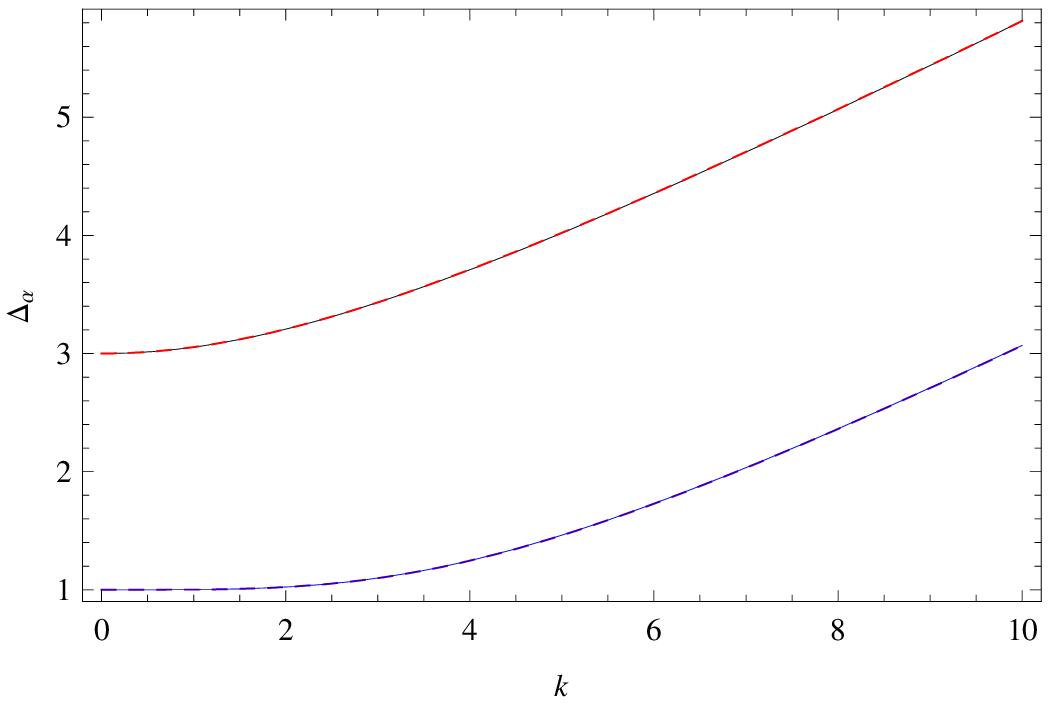}
\caption{
Left : Critical exponents in the presence of the CS term,
showing 
$\Delta_{\tilde{\kappa}}^{2} (\bm{k}) 
<\Delta_{\tilde{\kappa}}^{4} (\bm{k}) 
< \Delta_{\tilde{\kappa}}^{1} (\bm{k})
< \Delta_{\tilde{\kappa}}^{3} (\bm{k})$. 
The minimum value of $\Delta_{\tilde{\kappa}}^{2} (\bm{k}_{c})$ is 
associated with the instability to the non-uniform current ordering. 
Right : 
Critical exponents in the absence of the CS term, showing 
$\Delta_{\tilde{\kappa}=0}^{2}(\bm{k}) 
= \Delta_{\tilde{\kappa}=0}^{4} (\bm{k}) 
< \Delta_{\tilde{\kappa}=0}^{1} (\bm{k}) 
= \Delta_{\tilde{\kappa}=0}^{3}(\bm{k})$.  
} \label{fig1}
\end{figure}

\begin{figure}
\includegraphics[width=0.23\textwidth]{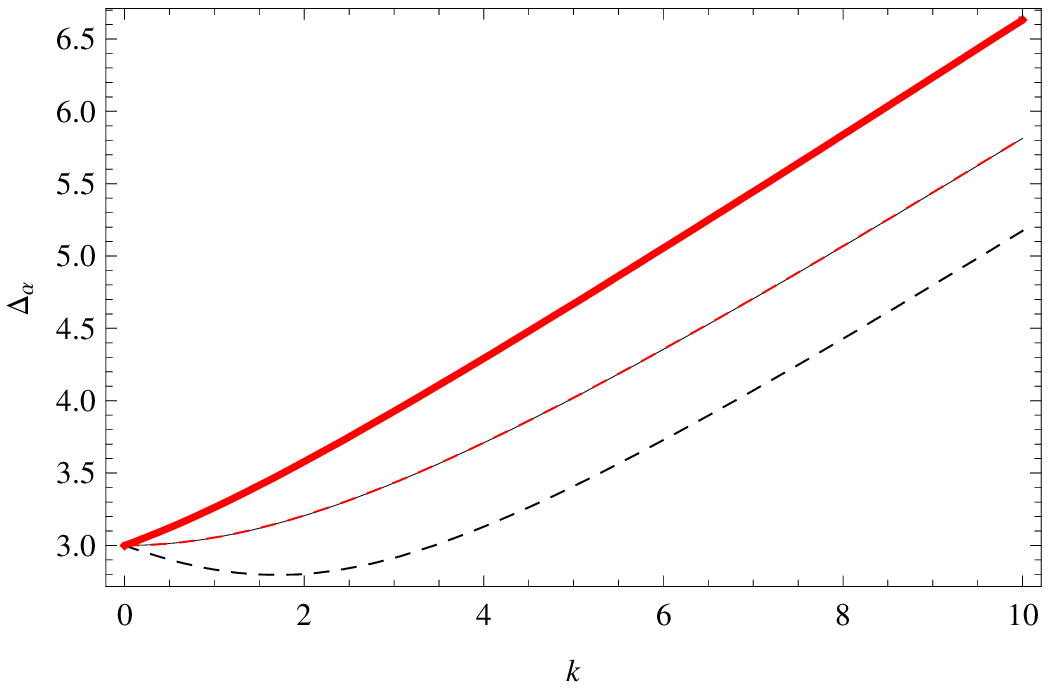}
\includegraphics[width=0.23\textwidth]{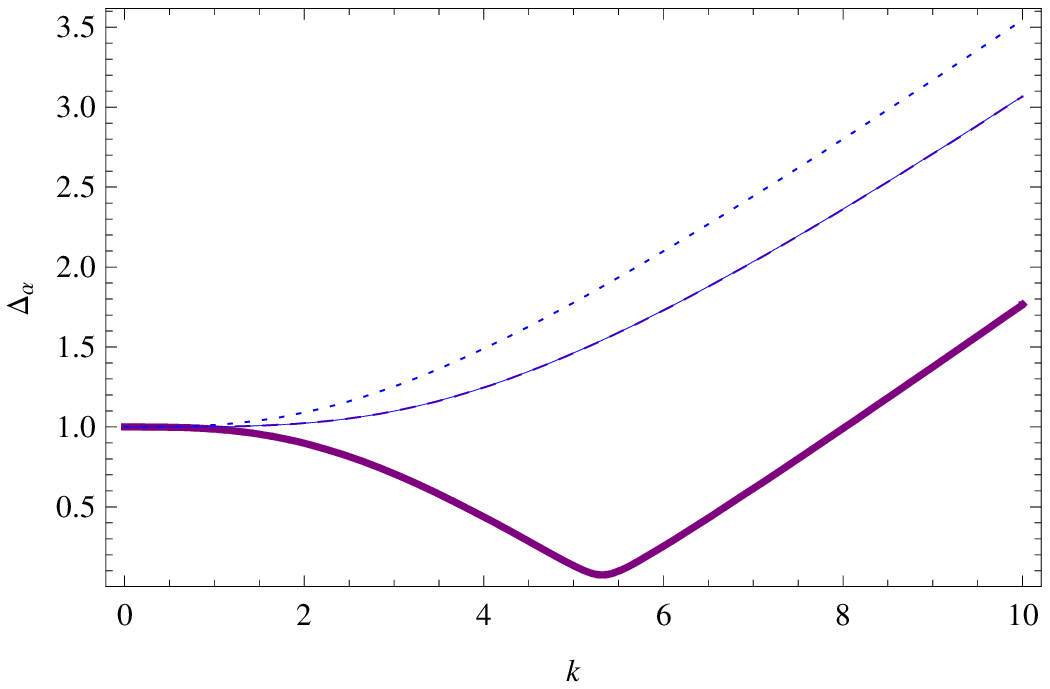}
\caption{
Left : $\Delta_{\tilde{\kappa}=0}^{1} (\bm{k}) 
= \Delta_{\tilde{\kappa}=0}^{3} (\bm{k})$ 
are split to 
$\Delta_{\tilde{\kappa}}^{1} (\bm{k}) 
< \Delta_{\tilde{\kappa}=0}^{1} (\bm{k}) 
=\Delta_{\tilde{\kappa}=0}^{3} (\bm{k}) 
< \Delta_{\tilde{\kappa}}^{3} (\bm{k})$. 
Right :
$\Delta_{\tilde{\kappa}=0}^{2} (\bm{k}) 
= \Delta_{\tilde{\kappa}=0}^{4} (\bm{k})$ 
are separated into 
$\Delta_{\tilde{\kappa}}^{2} (\bm{k}) 
< \Delta_{\tilde{\kappa}=0}^{2} (\bm{k}) =
\Delta_{\tilde{\kappa}=0}^{4} (\bm{k}) 
< \Delta_{\tilde{\kappa}}^{4} (\bm{k})$.
}
\label{fig2}
\end{figure}

We interpret the emergence of novel critical exponents as
$U(1)$ charge fractionalization. 
The $U(1)$ current with the conformal
dimension $\Delta_{\tilde{\kappa}=0}^{1} (\bm{k}_{c})$ becomes 
fractionalized into that with 
$\Delta_{\tilde{\kappa}}^{1} (\bm{k}_{c})$, where the latter
$U(1)$ current carries smaller $U(1)$ charge and has smaller conformal
dimension. 
Then, the other $U(1)$ current with
$\Delta_{\tilde{\kappa}}^{3} (\bm{k}_{c})$ may be identified 
with some composites of the $U(1)$ current with 
$\Delta_{\tilde{\kappa}}^{1}(\bm{k}_{c})$, 
where $\Delta_{\tilde{\kappa}}^{3} (\bm{k}_{c})$ should be
larger than $\Delta_{\tilde{\kappa}}^{1} (\bm{k}_{c})$, 
satisfied indeed. 
Essentially the same physics is also applied to
$\Delta_{\tilde{\kappa}}^{2} (\bm{k}_{c})$ and
$\Delta_{\tilde{\kappa}}^{4} (\bm{k}_{c})$.

It is important to understand the role of the AdS$_{2}$ geometry
at low energies. If we do not take the extremal limit of the
RN-AdS$_{5}$, the CS term does not modify any anomalous critical
exponents, where its role turns out to result in additional
analytic expansions for momentum and frequency \cite{mstt}.
However, the CS term in the emergent AdS$_{2}$ geometry generates
novel critical exponents, which will be associated with novel
excitations.

The AdS$_{2}$ geometry reminds us of the dynamical mean-field
theory (DMFT) framework \cite{DMFT} as the dual CFT, where the
extended DMFT of the Kondo lattice model gives rise to the local
critical theory \cite{EDMFT}. Recently, Sachdev proposed the
critical field theory of the disordered Kondo-Heisenberg model as
the CFT dual to the AdS$_{2}$ gravity theory with fermionic or
bosonic matters \cite{Sachdev_CFT}.
One of the authors has suggested that symmetry associated with
charge and spin fluctuations becomes enlarged at the local quantum
critical point and this emergent enhanced symmetry allows us to
assign a nontrivial quantum number to an instanton excitation
\cite{Kim_Disorder}. These topological excitations are identified
with spinons and holons, respectively. Unfortunately, it is not
clear at all how the symmetry enhancement appears to allow
topological excitations in the AdS$_{2}$ gravity theory
\cite{Sachdev_CFT}.

The present dual gravity theory does not have either fermionic or
bosonic matters. 
As discussed in the introduction, the corresponding conformal field 
theory will be given in terms of strongly coupled $U(1)$ charge currents 
with the associated chiral anomaly. 
In this respect the origin of the locality is not clearly figured out. 
Although the circulating current model~\cite{Varma_Current_Model}
attracts our interest, the connection with the AdS$_{2}$ dual gravity 
is not clear.

The interplay between interaction and $\theta$ vacua has been
discussed, based on the EM-CS theory in 
AdS$_{3}$~\cite{Jensen_AdS3}. 
The chiral anomaly for symmetry currents in
(1+1)D CFTs turns out to determine their correlators completely,
where a perfect metallic state appears to be distinguished from
the present RN-AdS$_{5}$ case, but consistent with the chiral
edge-state picture. 
On the other hand, the absence of the CS term, i.e.\ 
the Einstein-Maxwell (EM) theory is interpreted to be the presence 
of an external current coupled with
the EM gauge field, where the external current serves an
additional source for the Weyl anomaly. 

It will be interesting to investigate the Yang-Mills-Chern-Simons
theory on the RN-AdS$_{5}$ in the extremal limit, where the
corresponding conformal field theory is expected to be in terms of
strongly correlated nonabelian currents with the chiral anomaly.
Recently, the nonabelian gauge theory with the $\theta$ term has
been suggested for fractional magnetoelectric effect in
interacting topological insulators \cite{FTI_EFT_Theta}, where
dyon excitations carrying fractional electric charge are
responsible for such an effect. 
Although we cannot estimate the role of supersymmetry and 
emergent locality, we speculate that some types of quantum 
criticality in such topological insulators may be associated with 
a generalized framework of the present description.

\section{Summary}

The interplay between correlations and topological terms has
proposed novel quantum states of matter and emergent excitations
beyond our intuition. 
However, it was outside the field theoretical framework 
to incorporate both nonperturbative quantum effects reliably 
above one dimension. 
This motivated us to investigate the Einstein-Maxwell-Chern-Simons 
theory on Reissner-Nordstr\"om-AdS$_5$ background, 
where the corresponding field theory is
expected to be in terms of strongly interacting $U(1)$ charge
currents with anomalous ``chiral'' currents.

An important aspect was the interplay between the Chern-Simons
term and the emergent AdS$_{2}$ geometry of the extremal limit. If
we do not take the extremal limit, the Chern-Simons term itself
does not generate any anomalous critical exponents in
current-current correlation functions, where only analytic
expansions appear for momentum and frequency. If we do not
introduce the Chern-Simons term in the AdS$_{2}$ geometry,
critical exponents become rather trivial, where the
non-analyticity results from the emergent locality.

We interpret the emergence of novel critical exponents as a result
of complicated current-pattern fluctuations, driven by the
Chern-Simons term. 
The Chern-Simons term gives rise to couplings 
between $x$- and $y$-components of gauge fields with anomalous 
correlations. 
Such anomalous correlations are expected to cause nontrivial 
current-current interactions in different directions, 
which may be responsible for non-uniform current-loop
excitations. 
On the other hand, the absence of the Chern-Simons
term does not result in complicated current patterns, where both
$x$- and $y$- directional current fluctuations have the same
conformal dimensions.

\acknowledgments

KSK appreciates helpful conversations with S.-S. Lee, G. Baskaran,
Y. Kim, K.-Y. Kim and S.-J. Sin. 
TT thanks Y. Matsuo, S.-J. Sin, S. Takeuchi and C.-M. Yoo for discussion 
in the initial stage of the present work. 
KSK and TT acknowledge the Max Planck Society (MPG), 
the Korea Ministry of Education, Science and Technology (MEST), 
Gyeongsangbuk-Do and Pohang City for the support of the Independent 
Junior Research Group at the Asia Pacific Center for Theoretical 
Physics (APCTP). 
KSK was also supported by the National Research Foundation of 
Korea (NRF) grant funded by the Korea government 
(MEST) (No. 2011-0074542).

\appendix*

\section{Minkowskian correlators \\
in AdS/CFT correspondence}

We here briefly summarize the prescription for the Minkowskian correlator
in AdS/CFT correspondence.
We here follow the prescription proposed in~\cite{ss}.
We work on the following 5D background,
\begin{equation}
 (\dd s)^2 = g_{\mu\nu}\dd x^\mu\dd x^\nu + g_{uu}(\dd u)^2,
\end{equation}
where $x^\mu$ and $u$ are the 4D and radial coordinates,
respectively.
We refer the boundary at $u=0$ and the horizon at $u=1$.
Let us consider a solution of an equation of motion in
this 5D background.
Suppose a solution of an equation of motion is given by
\begin{equation}
\phi(u,x) =
\!\int\!\frac{\dd^4 k}{(2\pi)^4}\ \mbox{e}^{ikx}f_k(u)\phi^0(k),
\end{equation}
where $f_k(u)$ is normalized such that $f_k(0)=1$ at the boundary.
After putting the equation of
motion back into the action,
the on shell action might be reduced to surface terms
\begin{equation}
S[\phi^0]
=\!\int\!\frac{{\rm d}^4 k}{(2\pi)^4}
\phi^0(-k){\cal G}(k, u)\phi^0(k)
\bigg|_{u=1}^{u=0}.
\label{on_shell_action}
\end{equation}
Here, the function $\mathcal G(k,u)$ can be written
in terms of $f_{\pm k}(u)$ and $\partial_u f_{\pm k}(u)$.
Accommodating GKP-W
relation~\cite{AdS_CFT_Conjecture2} to Minkowski spacetime,
Son and Starinets proposed the formula to get the retarded correlation functions,
\begin{equation}
G(k)
=
2{\cal G}(k, u)
\bigg|_{u=0}. 
\label{green_function}
\end{equation}
We define the retarded correlation function we discuss in this paper: 
\begin{eqnarray}
&&
G_{\mu\ \nu}(\omega, k)
\nonumber 
\\
&&
\qquad =
-i\!\int\!\frac{\dd^2x}{(2\pi)^2}
\ {\rm e}^{-i\omega t+ikz}\theta(t)\langle[J_\mu(t, z), \ J_\nu(0, 0)]\rangle,
\nonumber
\\
&&G_{\mu\nu\ \rho\sigma}(\omega, k)
\nonumber 
\\
&&
\qquad
=
-i\!\int\!\frac{\dd^2x}{(2\pi)^2}
\ {\rm e}^{-i\omega t+ikz}\theta(t)
\langle[T_{\mu\nu}(t, z), \ T_{\rho\sigma}(0, 0)]\rangle,
\nonumber 
\\
&&G_{\mu\ \rho\sigma}(\omega, k)
\nonumber 
\\
&&
\qquad 
=
-i\!\int\!\frac{\dd^2x}{(2\pi)^2}
\ {\rm e}^{-i\omega t+ikz}\theta(t)
\langle[J_\mu(t, z), \ T_{\rho\sigma}(0, 0)]\rangle,
\nonumber 
\\
\end{eqnarray}
where the momentum in this paper is taken to $z$-direction,
and the operators $J_\mu(t, z)$ and $T_{\mu\nu}(t,z)$ 
are the $U(1)$ current and the energy-momentum tensor, 
respectively.

\end{document}